\documentclass[aps,pre,twocolumn,showpacs]{revtex4}
\usepackage{epsfig}
\usepackage{times}
\begin{document}

\title{Conditional strategies and the evolution of cooperation in spatial public goods games}

\author{Attila Szolnoki$^1$ and Matja{\v z} Perc$^2$}
\affiliation
{$^1$Research Institute for Technical Physics and Materials Science,
P.O. Box 49, H-1525 Budapest, Hungary \\
$^2$Faculty of Natural Sciences and Mathematics, University of Maribor, Koro{\v s}ka cesta 160, SI-2000 Maribor, Slovenia}

\begin{abstract}
The fact that individuals will most likely behave differently in different situations begets the introduction of conditional strategies. Inspired by this, we study the evolution of cooperation in the spatial public goods game, where besides unconditional cooperators and defectors, also different types of conditional cooperators compete for space. Conditional cooperators will contribute to the public good only if other players within the group are likely to cooperate as well, but will withhold their contribution otherwise. Depending on the number of other cooperators that are required to elicit cooperation of a conditional cooperator, the latter can be classified in as many types as there are players within each group. We find that the most cautious cooperators, such that require all other players within a group to be conditional cooperators, are the undisputed victors of the evolutionary process, even at very low synergy factors. We show that the remarkable promotion of cooperation is due primarily to the spontaneous emergence of quarantining of defectors, which become surrounded by conditional cooperators and are forced into isolated convex ``bubbles'' from where they are unable to exploit the public good. This phenomenon can be observed only in structured populations, thus adding to the relevance of pattern formation for the successful evolution of cooperation.
\end{abstract}

\pacs{89.75.Fb, 87.23.Ge, 87.23.Kg}
\maketitle

\section{Introduction}
The origins of prosocial behavior in groups of unrelated individuals are difficult to trace down. There exist ample evidence indicating that between-group conflicts may have been instrumental for enhancing in-group solidarity \cite{bowles_11}. On the other hand, some argue that our pre-human ancestors may have been confronted by more pressing challenges then simply to avoid being wiped out by their neighbors. About two million years ago some hominids were beginning to evolve larger brains and body size and to mature more slowly than other apes, which likely procreated serious challenges in rearing offspring that survived \cite{peters_83, calder_84}. Hence, alloparental care and provisioning for someone else's young have also been proposed as viable for igniting the evolution of the remarkable other-regarding abilities of the genus \textit{Homo} that we witness today \cite{hrdy_11}. Regardless of its origins, it is a fact that cooperation in groups is crucial for the remarkable evolutionary success of the human species, and it is therefore of the outmost importance to identify mechanisms that might have spurred its later development \cite{nowak_11}.

Evolutionary game theory \cite{sigmund_93, weibull_95, hofbauer_98, nowak_06, sigmund_10} is firmly established as the theoretical framework of choice for those studying the emergence and sustainability of cooperation at different levels of organization \cite{axelrod_84}. Recent reviews attest clearly to the fact that interdisciplinary approaches, linking together knowledge from biology, sociology, economics as well as mathematics and physics, are especially successful in identifying new ways by means of which the successful evolution of cooperation amongst selfish and unrelated individuals can be understood \cite{doebeli_el05, nowak_s06, szabo_pr07, schuster_jbp08, perc_bs10}. The public goods game, in particular, has proven itself times and again as the classic paradigm that succinctly captures the essential social dilemma that emerges as a consequence of group and individual interests being inherently different, which may ultimately result in the ``tragedy of the commons'' \cite{hardin_g_s68}. Governed by group interactions, the public goods game requires that players decide simultaneously whether they wish to contribute to the common pool, \textit{i.e.} to cooperate, or not. Regardless of the chosen strategy, each member of the group receives an equal share of the public good after the initial investments are multiplied by a synergy factor that takes into account the added value of collaborative efforts. Evidently, individuals are best off by not contributing anything to the common pool, \textit{i.e.} by defecting, while the group, and indeed the society as a whole, is most successful if everybody cooperates.

Recent research has made it clear that spatial structure plays a pivotal role by the evolution of cooperation, as comprehensively reviewed in \cite{szabo_pr07}. Inspired by the seminal paper introducing games on grids \cite{nowak_n92b}, evolutionary games on graphs and complex networks \cite{abramson_pre01, ebel_pre02, zimmermann_pre04, vukov_pre05, lieberman_n05, santos_pnas06, lozano_ploso08, santos_n08, szolnoki_epl09, liu_rr_pa10, poncela_njp09, zschaler_njp10, zhang_j_pa11, van-segbroeck_njp11, lee_s_prl11, poncela_pre11, dai_ql_njp10, wu_t_pone11, gomez-gardenes_c11} have proven instrumental in raising the awareness of the fact that relaxing the simplification of well-mixed interactions may lead to qualitatively different results that are due to pattern formation and intricate organization of the competing strategies, which reveals itself in most unexpected ways. Specifically for the spatial public goods game \cite{wakano_pnas09, szolnoki_pre09c}, it has recently been shown that inhomogeneous player activities \cite{guan_pre07}, appropriate partner selection \cite{wu_t_pre09, zhang_hf_epl11}, diversity \cite{yang_hx_pre09, ohdaira_acs11, perc_njp11}, the critical mass \cite{szolnoki_pre10}, heterogeneous wealth distributions \cite{wang_j_pre10b}, the introduction of punishment \cite{brandt_prsb03, helbing_ploscb10} and reward \cite{szolnoki_epl10}, as well as both the joker \cite{arenas_jtb11} and the Matthew effect \cite{perc_pre11}, can all substantially promote the evolution of public cooperation.

Apart from rare exceptions, the large majority of previously published works assumed unconditional strategies, \textit{i.e.} cooperators that always cooperated and defectors that always defected. Nevertheless, the usage of unconditional strategies constitutes a simplification that deserves further exploration. It is a fact that individuals, be it humans or animals, will likely behave differently under different circumstances. This invites the introduction of conditional strategies, by means of which such considerations can be appropriately taken into account. With this motivation, we here study the evolution of cooperation in the spatial public goods game containing conditional cooperators. Conditional cooperators will contribute to the common pool only if there is a sufficiently high number of other conditional cooperators in the group. If not, conditional cooperators will defect, at least until the group acquires more players that are likely to cooperate. While the details of the model and the main results will be presented in the following two sections, beforehand, the key finding of this work is that conditional cooperators are able to quarantine defectors into isolated convex ``bubbles'' from where they are unable to exploit the public good, and in doing so warrant completely defector-free states even if the synergy factor is close to one. Perhaps even more interestingly, we find that just the signalling of the willingness to cooperate, instead of a hard promise, is sufficient to elevate the level of collaborative efforts. As we will show, these observations rely on the spatial structure and cannot be observed in well-mixed systems, although they are robust against the topological variations of the interaction network and the group size.

\section{Spatial public goods game with conditional strategies}
The public goods game is staged on a square lattice with periodic boundary conditions where $L^2$ players are arranged into overlapping groups of size $G=5$ such that everyone is connected to its $G-1$ nearest neighbors. Accordingly, each individual belongs to $g=1,\ldots,G$ different groups. Initially each player on site $x$ is designated either as a conditional cooperator ($s_x = C_i$), where $i=0,\ldots,G-1$, or defector ($s_x = D$) with equal probability. Conditional cooperators contribute a fixed amount (here considered being equal to $1$ without loss of generality) to the public good only if there are at least $i$ other players within the group $g$ who are also willing to cooperate (whose strategy is $C_0$, $C_1$, $C_2$, $C_3$ or $C_4$) while defectors contribute nothing. Formally $C_0$ thus returns unconditional cooperators $C$ while $C_G$ returns unconditional defectors $D$. Note that in the presence of a player having strategy $s_x=C_G$ there cannot be $G$ other conditional cooperators within a group. The sum of all contributions in each group is multiplied by the synergy factor $r$ and the resulting public goods are distributed equally amongst all the group members irrespective of their contributions.

Monte Carlo simulations of the game are carried out comprising the following elementary steps. A randomly selected player $x$ plays the public goods game with its $G-1$ partners as a member of all the $g$ groups, whereby its overall payoff $P_{s_x}$ is thus the sum of all the payoffs acquired in the five groups. Next, player $x$ chooses one of its nearest neighbors at random, and the chosen co-player $y$ also acquires its payoff $P_{s_y}$ in the same way. Finally, player $x$ enforces its strategy $s_x$ onto player $y$ with a probability $w(s_x \to s_y)=1/\{1+\exp[(P_{s_y}-P_{s_x})/K]\}$, where $K=0.5$ quantifies the uncertainty by strategy adoptions \cite{szolnoki_pre09c}, implying that better performing players are readily adopted, although it is not impossible to adopt the strategy of a player performing worse. Such errors in decision making can be attributed to mistakes and external influences that adversely affect the evaluation of the opponent. Each Monte Carlo step (MCS) gives a chance for every player to enforce its strategy onto one of the neighbors once on average. The average densities of conditional cooperators ($\rho_{i}$) and defectors ($\rho_{D}$, alternatively denoted as $C_5$ and $\rho_5$) were determined in the stationary state after sufficiently long relaxation times. Depending on the actual conditions (proximity to phase transition points and the typical size of emerging spatial patterns) the linear system size was varied from $L=180$ to $720$ and the relaxation time was varied from $10^4$ to $10^6$ MCS to ensure proper accuracy.

It is worth pointing out that this is not a threshold-type model because the goods will always be shared between all the group members, even if the conditional cooperators will not all contribute. We note that the condition for conditional cooperators to cooperate introduced above is the most soft in terms of how many players are actually expected to cooperate. More precisely, it is likely that a very cautious cooperator (with a high $i$ value) will not cooperate, even though it may itself be reason enough for a less cautious cooperator to do so. Hence, in our model conditional cooperators require only a positive signal, or what can be interpreted as an ``easy promise'' from other group members, rather than a definite mutual agreement to contribute to the common pool. A much more stricter and sophisticated condition would be that a player having $s_x=C_i$ will cooperate only if there are at least $i$ other players in the group whose index is less or equal to $i$. This rule, imposing thus much stricter conditions, can only be applied if there is also at least one $C_G$ player (unconditional defector) in the group. Note that without it this definition yields misleading commands to conditional defectors. For example, in a group containing players $C_0$, $C_3$, $C_3$, $C_4$ and $C_4$, the strict $C_i \leq C_j$ rule would dictate defection for all $C_3$ players. In the following, we will refer to the dynamics relying on the ``more careful'' conditional strategies as the strict rule. We will comment on the outcome of such and other alternative models in the next section, where we now proceed with presenting the main results.

\section{Results}

\begin{figure}
\centerline{\epsfig{file=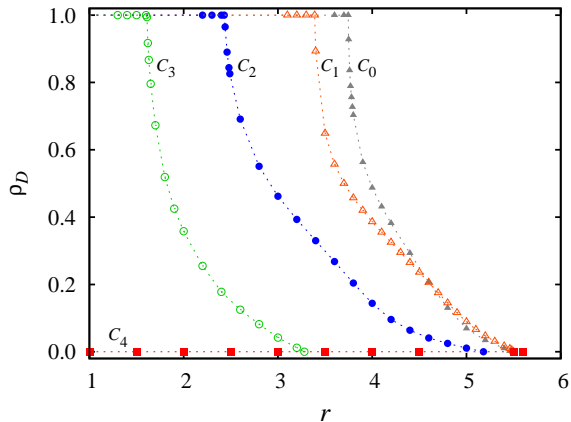,width=8cm}}
\caption{\label{three} (Color online) The fraction of defectors $\rho_D$ as a function of the synergy factor $r$, as obtained for different combinations of strategies that compete for space on the square lattice. Besides pure cooperators ($C$) and defectors ($D$), $1/3$ of the lattice is initially occupied by one conditionally cooperative strategy ($C_i$), as marked by each depicted curve. It can be observed that the higher the value of $i$, the earlier (at lower $r$) the downfall of $\rho_D$. Remarkably, if the most cautious conditional cooperators are introduced ($C_4$), defectors are completely defeated irrespective of $r$. We note that the results obtained with the related 3-strategy strict rule model are identical.}
\end{figure}

It is instructive to first examine the evolution of a subset of all the possible strategies. Figure~\ref{three} shows the outcomes of three-strategy games, where besides the unconditional cooperators $C_0$ and defectors $C_G=D$ one conditionally cooperative strategy ($C_1$, $C_2$, $C_3$ or $C_4$) is initially present. Depicted is the stationary density of defectors $\rho_D$ versus the synergy factor $r$ for the four possible strategy triples, as well as for the traditional two-strategy version of the spatial public goods game ($C_0$ curve). In the latter case, cooperators die out at $r \leq 3.748$, which is a well-known result \cite{szolnoki_pre09c}. Additionally introducing one type of conditional cooperators to the traditional setup continuously decreases the minimally required $r$ for cooperative behavior to survive as $i$ increases. Most remarkably, if initially unconditional cooperators $C_0$ and defectors $C_G=D$ and conditional cooperators $C_4$ each occupy $1/3$ of the lattice, we find that defectors cannot survive even in the $r \to 1$ limit ($C_4$ curve). This indicates that simple conditional strategies have ample potential for elegantly avoiding the ``tragedy of the commons'' even under the worst of conditions.

\begin{figure}[b]
\centerline{\epsfig{file=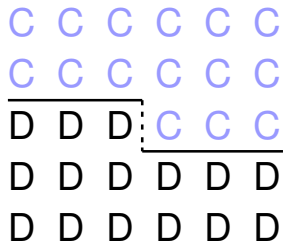,width=4cm}}
\caption{\label{stability} (Color online) Schematic presentation supporting the interface stability analysis of competing domains. The leading process, which modifies the interface between the ordered domains more intensively, is the invasion across the border marked by the dashed line.}
\end{figure}

The critical value of $r$ where cooperators die out can be estimated by means of a simple approach that considers the competition between two ordered domains of strategies \cite{szabo_jtb12}. As Fig.~\ref{stability} illustrates, the elementary change which modifies the step-like interface between competing domains is an invasion across the dashed line between unequal strategies. Assuming $C_j$ players as conditional cooperators and unconditional defectors, the accumulated payoffs of competing players are as follows:
\begin{equation}
{r \over G} \sum_{i=j+1}^{G-1} i = {r \over G} \sum_{i=j+1}^{G} i - (G-j) \,\,.
\label{equal}
\end{equation}
From this equation the critical synergy factor for the conditional cooperator strategy $C_j$ is $r_c^j = G-j$. Thus, even this simple analysis is able to reproduce the decreasing critical values of $r_c$ by increasing $j$, and moreover, warns that unconditional defectors cannot exist if a $C_{G-1}$ strategy is present. Evidently, other elementary processes are also possible, but their contributions to the boundary velocity are smaller, and to consider them would make this analysis untraceable. What is important is to note that the result is independent of the group size $G$, and is qualitatively valid for all lattice types. This is a straightforward consequence of the multi-point interaction of public goods games, which diminishes several microscopic differences of graphs and makes topological features like the clustering coefficient irrelevant \cite{szolnoki_pre09c}. The latter are, of course, essential for games that are governed by pair-wise interactions \cite{szabo_pr07}. As evidenced by the stability analysis, the crucial property in the presently studied model, however, is the ``spatiality'', allowing interfaces that separate domains of different strategies, which can only be fulfilled in structured populations.

\begin{figure}
\centerline{\epsfig{file=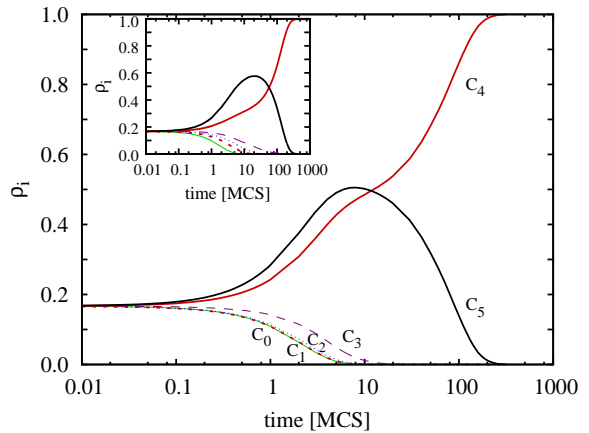,width=8cm}}
\caption{\label{lowr} (Color online) Time evolution of the complete six-strategy public goods game with unconditional cooperators ($C_0$) and defectors ($C_5$) as well as the four conditionally cooperative strategies ($C_1$, $C_2$, $C_3$ and $C_4$), as obtained for $r=1.05$. As can be deduced from results presented in Fig.~\ref{three}, at such low synergy factors all but the $C_4$ strategy are outperformed by defectors. However, after the defector-induced extinction of $C_{0\ldots3}$, the most cautious conditional cooperators ($C_4$) are able to completely invade the defectors ($C_5$). Inset shows the same evolution as obtained if using the strict rule model, and it can be observed that the final outcome is the same.}
\end{figure}

Turning to the complete six-strategy version of the spatial public goods game, we find that our main conclusion, arrived at based on the analysis of different three-strategy games, remains fully valid. In Fig.~\ref{lowr}, we first present characteristic time courses of all six strategies over time as obtained for a very low value of $r$. As expected based on results presented in Fig.~\ref{three}, unconditional cooperators $C_0$, as well conditional cooperators $C_1$, $C_2$ and $C_3$, albeit marginally latter, all die out fast due to invading defectors $C_5$. However, after the defectors are left on their own with $C_4$, they cannot withstand the invasion of this strain of conditional cooperators and die out. Interestingly, although $C_4$ appear to be eating into the territory of defectors from the very beginning of the evolutionary process, their full potential is unleashed only after all the other ``less cautious'' conditional cooperators die out. This is because although $C_4$ are obviously impervious to defectors, this is not necessarily the case with regards to other conditionally cooperative strategies. In particular, it may well be that under certain circumstances the lesser criteria for when to cooperate may yield a temporary advantage of $C_{0\ldots3}$ over $C_4$. Thus in fact, the other cooperative strategies hinder $C_4$ at effectively invading defectors by invading $C_4$ themselves. This, however, is very short lived as defectors are able to invade $C_{0\ldots3}$ extremely effectively at $r \to 1$. In a rather twisted turn of events, defectors, by invading $C_{0\ldots3}$, actually pave the way themselves towards a premature extinction. Although one could thus, in principle a least, hypothesize an alliance between $C_{0\ldots3}$ and $C_5=D$ to successfully invade $C_4$, our simulations reveal that the evolutionary window for such a complicated alliance to remain stable is too small to exist.

\begin{figure}
\centerline{\epsfig{file=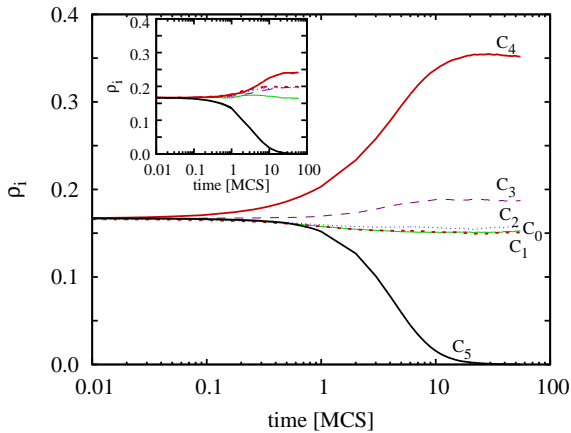,width=8cm}}
\caption{\label{highr} (Color online) Time evolution of the complete six-strategy public goods game with unconditional cooperators ($C_0$) and defectors ($C_5$) as well as the four conditionally cooperative strategies ($C_1$, $C_2$, $C_3$ and $C_4$), as obtained for $r=4.5$. At such a high synergy factor all five cooperative strategies $C_{0\ldots4}$ are able to withstand being wiped out by defectors. In fact, the latter are forced to extinction primarily by $C_4$, and to a much lesser extend by $C_3$. As soon as defectors die out, however, all cooperative strategies become equivalent, and their evolution becomes identical to that of the voter model. As in Fig.~\ref{lowr}, the inset shows the same outcome for the strict rule model.}
\end{figure}

At high synergy factors, however, the outcome of the six-strategy public goods game is significantly different. As evidenced by results presented in Fig.~\ref{highr}, at $r=4.5$ all the cooperative strategies are able to withstand being invaded by defectors. In fact, both $C_4$ and to a substantially lesser degree $C_3$ are able to do the exact opposite, which is to gain ground on the expense of retreating $C_5=D$. Importantly, however, after the defectors die out all five remaining strategies $C_{0\ldots4}$ become completely equivalent. Note that due to the soft condition, requiring only the presence of a certain number of conditional cooperator within the group, but regardless of their type (see Section II for detail), all group members now receive the required number of positive signals from others to actually go ahead and contribute to the common pool. Henceforth, the evolution becomes identical to that of the voter model \cite{cox_ap83}, entailing logarithmically slow coarsening in the absence of surface tension \cite{dornic_prl01}. The final stationary state is thus determined primarily by the share of the square lattice that is occupied by any given strategy at the time of defector extinction.

\begin{figure}
\centerline{\epsfig{file=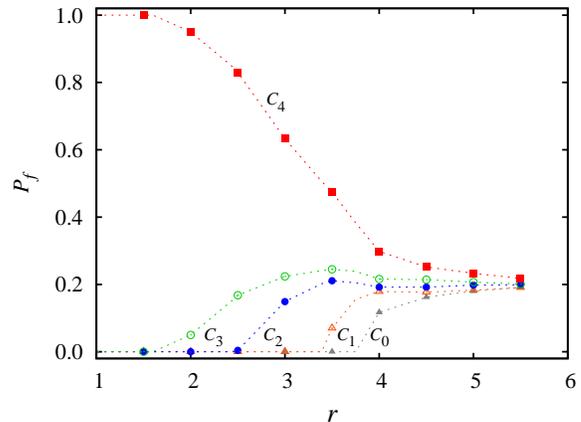,width=8cm}}
\caption{\label{fix} (Color online) Fixation probabilities $P_f$ of the complete six-strategy game to eventually arrive at a pure $C_i$ phase in dependence on the synergy factor $r$. It can be observed that while at $r \to 1$ the fixation at $C_4$ is practically unavoidable, in the high $r$ limit all five cooperative strategies ($C_{0\ldots4}$) become equally probable as the victors of the evolutionary process. Importantly, regardless of $r$ defectors are unable to survive, let alone dominate.}
\end{figure}

This interpretation can be made more precise by determining the fixation probabilities of the cooperative strategies in dependence on $r$. Results presented in Fig.~\ref{fix} indicate that, because of the neutral relations between the five cooperative strategies, which set in after the defector die out, the governing voter-model dynamics will, through coarsening, result in a homogeneous state where the system fixates into one of the remaining $C_i$ strategies, where $i<G$. The fixation probability depends on the fraction of competing strategies at the time of $\rho_D\to0$, which in turn depend on the effectiveness of the cooperative strategies to invade unconditional defectors, and to a lesser degree also on their effectiveness to invade each other. Based on the time evolutions presented in Figs.~\ref{lowr} and \ref{highr}, it is understandable that at low values of $r$ the fixation probability of $C_4$ will be practically one, while in the opposite limit the eventual dominance of either cooperative strain will be equally probable (see Fig.~\ref{fix}).

\begin{figure}
\centerline{\epsfig{file=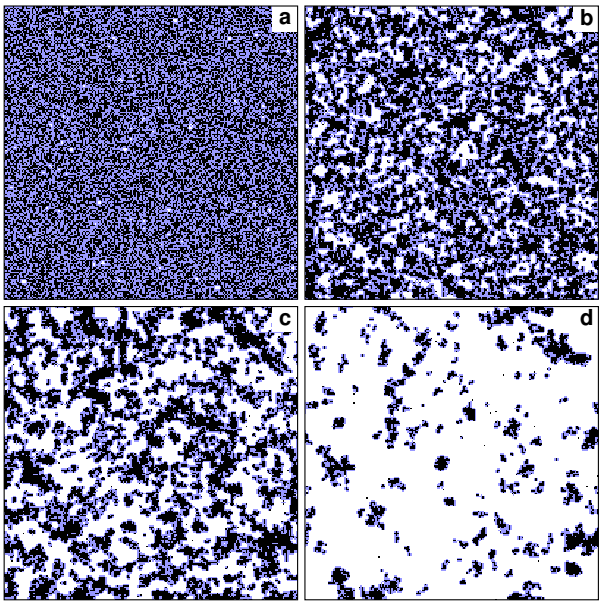,width=8cm}}
\caption{\label{snaps} (Color online) Characteristic snapshots depicting the competition between conditional cooperators $C_4$ and unconditional defectors $C_5=D$ when starting from a random initial state, as obtained for $r=1.05$. Black are $C_5$, white are $C_4$ when they cooperate in at least three out of five groups, while light blue (gray if printed BW) are $C_4$ that cooperate in less than three of the five groups where they are members. The snapshots were taken at $0$ (a), $10$ (b), $30$ (c) and $100$ (d) full Monte Carlo steps (MCS). The final state is a pure $C_4$ phase (all players depicted white), which is not shown. It can be observed that initially all the cooperators are practically inactive or hidden [light blue and black dominates in panel (a)]. Only after the spatial reciprocity takes effect and first cooperative clusters are formed do the conditional cooperators actually start cooperating, although they do so only in the interior of the clusters [see panel (b)]. At the borders separating the two competing strategies, however, virtually all $C_4$ remain inactive as they are unable to gather the required number of positive signals from their neighbors [see panel (c)]. This thin intermediate layer of hidden and inactive $C_4$ then acts as a shield that makes it incredibly difficult for defectors to invade. In fact, defectors become effectively quarantined into ``bubbles'' from where they are unable to exploit cooperators [see panel (d) and Fig.~\ref{bubble} for a zoom-in]. Ultimately, this mechanism results in the ``tragedy of the defectors'' irrespective of the value of $r$. The linear system size used here is $L=200$.}
\end{figure}

\begin{figure}
\centerline{\epsfig{file=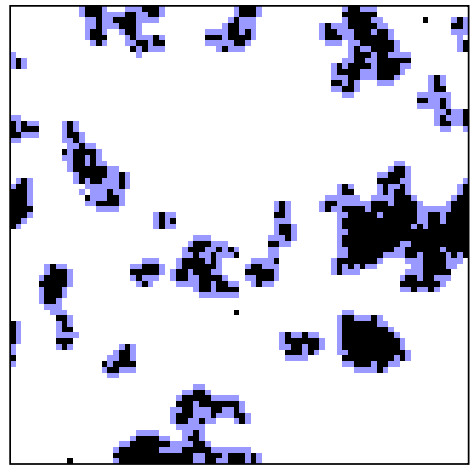,width=6.25cm}}
\caption{\label{bubble} (Color online) An $80 \times 80$ zoom-in of Fig.~\ref{snaps}(d), demonstrating clearly the spontaneous emergence of convex isolated ``bubbles'' of defectors (depicted black) that are contained by inactive conditional cooperators of type $C_4$ (depicted light blue, gray if printed BW). While the latter will predominantly cooperate with the bulk of active conditional cooperators of the same type (depicted white), they will certainly defect in the opposite direction where there are unconditional defectors. Consequently, defectors cannot exploit $C_4$ type players, which leads to a gradual but unavoidable shrinkage of the defector quarantines.}
\end{figure}

In order to reveal the main mechanism behind the rather remarkable inability of defectors to survive in the presence of $C_4$, it is instructive to visualize the spatial patterns emerging as a consequence of their direct competition. Figure~\ref{snaps} features a series of four characteristic snapshots that were taken at different times [increasing from panel (a) to panel (d)] where only the two mentioned strategies compete for space. While defectors are depicted black, for convenience we graphically distinguish between two types of $C_4$ players. Namely between such that are predominantly active as cooperators, and such that are predominantly inactive or hidden. The criterion separating the two is simply the number of groups in which the player has actually contributed to the common pool. If the player, at the time the snapshot was taken, has cooperated in three or more out of the five groups where it is member, we mark it as active and depict it white, while otherwise, if it has cooperated in two or fewer groups, we mark it as inactive and depict it light blue (gray if printed BW). With this graphical distinction, we reveal that the reason why defectors cannot exploit $C_4$ effectively is due to the spontaneous emergence of very persistent interfaces of inactive $C_4$ players that separate cooperative and defective domains. Since $C_4$ players immediately stop cooperating in a group that contains at least one $C_5=D$, the defectors cannot collect large, competitive payoffs near the interfaces (and certainly not in the middle of the sea of $D$). On the other hand, hidden cooperators are still capable to collect significant payoffs from $C_4$ players that are on the opposite side of the interface, where in general the condition to actually cooperate will be fulfilled. In this way, hidden cooperators don't just shield the active cooperators from the invasion of defectors, but they can also effectively invade defectors to eventually completely dominate the whole population. The phalanx of hidden cooperators will quarantine unconditional defectors into convex isolated ``bubbles'', as demonstrated in Fig.~\ref{bubble}, which ultimately leads to an unavoidable ``tragedy of the defectors''.

From the described workings of the mechanism, it is clear that it cannot emerge under well-mixed conditions, as then players adopting the $C_4$ strategy will essentially never actually cooperate, given that an encounter with at least one defector is virtually unavoidable. Despite of this fact $C_4$s can survive, but they will always reveal only their defector face. Accordingly, the ``tragedy of the commons'' cannot be avoided by means of similar conditional strategies in well-mixed settings of the public goods game. On the other hand, it is also clear that the mechanism is robust and potent not only on the square lattice, but in fact on all other types of interaction graphs where long-standing bonds between players are assumed (note that certain coevolutionary rules \cite{perc_bs10}, especially such that rely on frequent rewiring of the links between players, may render the mechanism dysfunctional). Finally, we emphasize another positive message of this study, which is that cooperation can be promoted simply by the signaling of others that they are willing to cooperate, rather than a firm oath that they will actually do so. We have observed that our results remain valid also when introducing the more sophisticated conditional strategies, as discussed in Section II, although we find the usage of the more elegant and simple model much more rewarding and interesting.

\section{Summary}
In summary, we have shown that an intuitive introduction of conditional cooperative strategies provides the ultimate boost to the mechanism of spatial reciprocity \cite{nowak_n92b}. In particular, the most cautious conditional cooperators provide an escape hatch out of the ``tragedy of the commons'' for all values of the synergy factor $r$ by spontaneously forming a protective shield between them and the defectors. The shield, however, makes it not only extremely difficult for defectors to exploit the collaborative efforts of others, but at the same time provides an evolutionary advantage to cooperators that enables their invasion of the territory of defectors, eventually leading to their complete dominance. The quarantining of defectors is crucial especially at very low values of $r$, where otherwise they can reap huge benefits on the expense of cooperators. At intermediate and high values of $r$, however, all the different strains of conditional cooperators become more and more able to withstand being wiped out by defectors on their own. Thus, as soon as defectors die out, the evolution of the remaining cooperative strategies become neutral and proceeds by means of coarsening that is characteristic for the voter-model-type dynamics \cite{dornic_prl01}. By determining the fixation probabilities in dependence on the synergy factor $r$, we have shown that in the low $r$ limit the fixation at $C_4$ (the most cautious conditional cooperators) is practically unavoidable, while in the high $r$ limit all five cooperative strategies ($C_{0\ldots4}$) become equally probable to emerge as the dominant trait. Regardless of $r$, however, the defectors are unable to survive the evolutionary process, which is a very rewarding discovery to arrive at simply by means of a conditional strategy ($C_4$). Conceptually at least, our approach can be related to a recent study by Vukov et al. \cite{vukov_jtb11}, where directed investments were introduced to the public goods game. In their model, however, a cooperator will necessarily invest somewhere, while in our case cooperators may remain dormant for long periods of time before eventually deciding to contribute to the common pool. In terms of potential implication of our findings, apart from their relevance for the successful evolution of prosocial behavior between selfish and unrelated individuals, from the biological point of view, the way inactive cooperators quarantine defectors and force them into convex isolated ``bubbles'' bears resemblance to the way the immune system works when trying to contain an infection \cite{szabo_jtb07}. We hope that this study will inspire future research aimed at investigating the role of conditional strategies in structured populations.

\begin{acknowledgments}
This research was supported by the Hungarian National Research Fund (grant K-73449) and the Slovenian Research Agency (grant J1-4055).
\end{acknowledgments}

\end{document}